\documentclass[12pt]{iopart}
\expandafter\let\csname equation*\endcsname\relax
\expandafter\let\csname endequation*\endcsname\relax
\usepackage{amssymb}
\usepackage{amsmath}
\usepackage{amsthm}
\usepackage[numbers,sort&compress,square]{natbib}
\usepackage{bm}
\usepackage{graphicx}
\usepackage{calc}
\usepackage[dvipsnames]{xcolor}
\usepackage{hyperref}
\usepackage[capitalise]{cleveref}
\usepackage{tabularray}

\theoremstyle{definition}

\theoremstyle{remark}

\newcommand{\mean}[1]{\left\langle#1\right\rangle}
\newcommand{\eexp}[1]{e^{#1}}
\newcommand{\norm}[1]{\left\lVert#1\right\rVert}

\begin{document}

\title{Binary Option Market Manipulation by Influencing Belief Dynamics}

\author[inst1]{Henry Waldhausen$^1$ and Christopher Griffin$^2$}
\address{
	$^1$Eberly College of Science,
	The Pennsylvania State University,
    University Park, PA 16802
    }
\address{
	$^2$Applied Research Laboratory,
	The Pennsylvania State University,
    University Park, PA 16802
    }
\eads{hww5202@psu.edu, griffinch@psu.edu}

\date{\today~-~Preprint}

\begin{abstract} Using techniques from information geometry, we construct a semi-Hamiltonian system modelling trader beliefs in a binary asset market and study the impact of inequality or asymmetry in beliefs, information, and power on price dynamics. We show that in a market with no inequality and $N$ completely symmetric traders, the resulting dynamics evolve on a $2N + 1$ dimensional manifold consisting of a $2N-2$ dimensional centre manifold, a $2$ dimensional stable manifold and a $1$ dimensional slow manifold. Introducing asymmetry into the traders has the potential to decrease the dimension of the centre manifold, which we prove using a parameter analysis. Using the belief model, we also study the impact of inter-agent communication, exogenous information and asymmetric purchasing power on price dynamics, showing that market bubbles can emerge when powerful traders produce outsize influence in the market, thus impacting other traders' beliefs as well as the price. This process is exacerbated when back-channel communication is permitted. The impact of areas of high curvature in belief space is also discussed. 
\end{abstract}

% Uncomment for keywords
\vspace{2pc}

\maketitle

\section{Introduction}
There have been several theoretical and simulation-based studies of markets within the econophysics literature (see \cite{M00,CS01,BVS06,SZSL07,S01,SP08,RP03,MZ08,CS01a,HP02,TFM18,D21} for examples) and the potential contributions from physics in understanding market dynamics are still being realised. Recent work by Zhou et al. \cite{ZZL22}  considers information delay in market models, while Dicks et al. consider learning by market participants \cite{DPG24}. Learning is also considered by Vives, who investigates the impact of market microstructure on learning \cite{V10}. Spin systems using Potts models \cite{B22} are studied by Bornholdt, while Lan and Fang \cite{LF24} use an Ising spin model for market dynamics, as do Krause and Bornholdt \cite{KB13}. Slightly older work by Zhang and Huang \cite{ZH10} and Cotfas \cite{C13} propose quantum mechanical explanations for chaotic market behaviour. We note this is just a sample of the much wider literature on this topic.

Surprisingly, inequality in markets, i.e., markets with groups of non-identical traders, are less frequently studied with most studies focusing on the relationship between societal inequalities and markets. Favilukis takes this approach in \cite{F13a}. Similarly, Melcangi and Sterk \cite{MS24} study dynamics (and inequality) in stock market participation and its resulting effects on monetary policy, rather than on the market itself. One area where ``fairness'' (or inequity vs. inequality) is seriously considered in financial markets is in the case of high-frequency trading (HFT). Angel and McCabe \cite{AM13} consider this, showing that there is no empiric measure by which HFT is fair or unfair to all market participants. Fishman and Hagerty \cite{FH92} consider insider trading, as a form of extreme information inequality and show that insider traders can lead to less efficient markets, despite providing information into them (through their trading actions). Perhaps closest in spirit to this paper is the work of Raberto et al. \cite{RCFM03} who investigate the impact of various trading strategies within the synthetic Genoa Artificial Stock Market, but do not consider the impact of traders with (e.g.) unequal capital.

In this paper, we study the impact of trader inequality on market behaviour, both theoretically and empirically. We do so using a simple model of a binary option market. Binary option markets \cite{BL78} are particularly simple market structures that are now banned in most regions because they (essentially) represent a naked bet on an outcome and are frequently the target of fraud \cite{T14}. Interestingly, they had been used on Wall Street during the 1800s \cite{RS04}. Prediction markets are common and permissible binary (or more generally $n$-ary) markets in which assets corresponding to future events (e.g., elections \cite{BFR97}, sports outcomes \cite{TZ88}, etc.) can be bought and sold, causing changes in the underlying asset prices. These markets were first studied by Hanson \cite{H90,H91,R97}. Since this initial work, they have been studied extensively \cite{WZ04,SWPG04,M06,BR03,WZ06,DJK20,CD16}. Chen et al. show that if a prediction market has a cost function with bounded loss, then it has an interpretation as a no-regret learning algorithm \cite{CV10}, thus relating prediction market dynamics to learning. For a comprehensive discussion of prediction markets, see Tziralis and Tatsiopoulos who provide a survey of work in this area through 2007 \cite{TT07}.

In the case of a binary (or $n$-ary market), asset prices can be interpreted as categorical probabilities \cite{M06,WZ06}, which allows one to re-interpret the market as a dynamic process on the Riemannian (information geometric) manifold defined by the Fisher information metric \cite{N20} using techniques from information geometry \cite{A16} and non-Euclidean dynamics \cite{MR13}. In this paper, we use results from Goehle and Griffin \cite{GG24} to model an individual trader's belief as a Bernoulli variable that likewise evolves on an information geometric manifold in response to observation of the price and (in some cases) external stimuli, such as the direct opinions of other traders. We show that for $N > 1$ identical traders the dynamics evolve on a $2N-2$ dimensional centre manifold, a $2$ dimensional stable manifold and a $1$ dimensional slow manifold. We provide sufficient conditions for asymmetry (inequality) in one of the traders to decrease the dimensionality of the centre manifold by two, and show the effect. We then study how other inequalities in the traders can lead to market instabilities, including potential belief bubbles that can cause purchase cascades. The impact of back-channel contact between traders is also studied in this context.

%%This is better in the introduction.
%We're using a model of belief in a binary options market to observe what happens to a stock price given certain initial conditions, external information, and a communication network among agents. Our model uses the Kullbach-Liebler Divergence to model the energy, or work, required to change an agent's belief about whether or not a stock should be bought, and the Barabasi-Albert Graph distribution to describe the communication network.

The remainder of this paper is organised as follows: We introduce background information and the generic model in \cref{sec:BackgroundModel}. Theoretical results on the models are derived in \cref{sec:TheoreticalResults}. The impact of communication, external information and asymmetry in trader power is studied in \cref{sec:ExternalInformation}. Theoretical results on market manipulation are presented in \cref{sec:Manipulation}. We present conclusions and future directions in \cref{sec:Conclusions}.

\section{Mathematical Background and Model}\label{sec:BackgroundModel}
%%Information Geometry

\subsection{Market Model}
Following \cite{GG23} we use a logarithmic market scoring rule (LMSR) to model a binary option market. In this setting, quantities of two assets are purchased by agents interacting in the market. Let $q_j$ ($j \in \{0,1\}$) denote the net quantities of the two assets purchased by all traders. Using a LMSR, the instantaneous spot price of each asset is,
\begin{equation*}
    p_j = \frac{\eexp{\beta q_j}}{\eexp{\beta q_0} + \eexp{\beta q_1}}.
\end{equation*}
The quantity $\beta$ is a liquidity constant that determines the speed of price change as a function of asset purchases. We assume $0 < \beta  < 1$ so that incremental purchases do not affect the price dramatically. %As we will see, a small $\beta$ introduces a natural delay in the correlation of the mean agent belief and the spot price, which is discussed in \ref{app:Delay}. 

We note in this form, the asset prices are defined by a two state Boltzmann distribution, which is equivalent to a Bernoulli distribution. Using the fact that $p_j = (1-p_{1-j})$, we can consider a single price $p$ (for asset class 1) and straightforward differentiation shows that,
\begin{equation}
    \dot{p} = \beta p (1-p)\left(\dot{q}_1 - \dot{q}_0\right).
    \label{eqn:Rawqdot}
\end{equation}
These are the dynamics of the parameter of the Bernoulli distribution defining asset prices, and we see that specifying the rate of change of the quantities then determines the dynamics of the entire system. 

Gampe and Griffin \cite{GG23} define $\dot{q}_1 - \dot{q}_0$ using a step-function, assuming assets must be purchased in whole quantities and based on a purchasing rule that uses information from a common external information source. In this paper, we assume that each agent in the market has a time-varying belief about the price, $\rho_i \in (0,1)$, where $i \in \{1,\dots,N\}$, and that agents can purchase fractional asset quantities so that,
\begin{equation*}
    \dot{q}_1 - \dot{q}_0 = \sum_i Q_i(\rho_i - p),
\end{equation*}
where $Q_i$ is the purchasing power of agent $i$ and agent $i$ will instantaneously purchase a fraction of asset $1$ if $\rho_i > p$. Otherwise, agent $i$ buys a fraction of asset $0$, which is equivalent to selling a fraction of asset $1$. The resulting price dynamics are then,
\begin{equation}
    \dot{p}=\beta p(1-p)\sum_i{Q_i\left(\rho_i-p\right)}.
    \label{eqn:pdot}
\end{equation} 

\subsection{Information Geometry}
To fully specify the dynamics of the market, we must specify the dynamics of $\rho_i$, agent $i$'s belief about the correct price. We do so using the Bernoulli belief model of Goehle and Griffin \cite{GG24,GG24a}, built on Friston's free energy principle \cite{FKH06,F09,F10,FDSH23}. Informally, this model assumes that Bayesian brains build and update models in an attempt to minimize ``surprise''. Following \cite{FDSH23}, we define ``surprise'' in terms of (free) entropy, implying that the free energy principle depends on Landauer's principle \cite{L61} (in some sense) relating energy used to information changed, in this case measured by free entropy. We provide the general information geometric framework below and then specialize to the Bernoulli belief model.

Following \cite{GG24,GG24a}, a belief is simply an appropriately defined time-varying probability distribution $p(\mathbf{x} | \bm{\eta})$ with parameters $\bm{\eta}$. We use the machinery of information geometry \cite{nielsen2018,N20,A16} to construct the required free entropy as well as an appropriate Riemannian manifold on which to define belief evolution. In the fully abstract setting, the free entropy (surprise) produced by the changing set of distribution parameters $\bm{\eta}$ in $p(\mathbf{x}|\bm{\eta})$ can be measured with the Fisher metric, whose quadratic form is given by,
\begin{equation*}
g_{jk}(\bm{\eta}) = \int_X \frac{\partial \log\left[p(\mathbf{x}|\bm{\eta})\right]}{\partial \eta_j}\frac{\partial \log\left[p(\mathbf{x}|\bm{\eta})\right]}{\partial \eta_k}p(\mathbf{x}|\bm{\eta})\, dx.
\end{equation*}
Using the Einstein summation convention, the kinetic energy action,
\begin{equation*}
    A = \frac{1}{2} \int_{\lambda_0}^{\lambda_f}  \mathbf{g}_{jk}(\bm{\eta}) \frac{\partial\eta^j}{\partial \lambda} \frac{\partial\eta^k}{\partial \lambda} \,d\lambda,
\end{equation*}
gives the square distance measured in nats along a parametrically defined path $\bm{\eta}(\lambda)$ with $\lambda \in [\lambda_0,\lambda_f]$ \cite{A16,N20}. This is the free entropy of the path. Here $\lambda$ is as a pseudo-time parameter, however in what follows it will be replaced by time in the usual sense. The free entropy measures the ``surprise'' (or information) needed to move  continuously from one belief to another along a given path in distribution space. 

The free entropy is intimately related to the Kullback-Liebler (KL) divergence between two probability distributions, $p(\mathbf{x}\vert\bm{\eta})$ and $q(\mathbf{x}\vert\bm{\eta})$, given by,
\begin{equation}
D_{KL}(p\vert q) = \int_X p(\mathbf{x}\vert\bm{\eta}) \log\left(\frac{p(\mathbf{x}\vert\bm{\eta})}{q(\mathbf{x}\vert\bm{\eta})}\right) d\mathbf{x}.
\end{equation}
In classical information theory, the KL divervence measures the average number of additional bits needed to encode data generated with distribution $p$ but assuming (a code for) distribution $q$. From the point of view of the Bayesian brain hypothesis, we can think of it as the amount of extra information needed to process an external belief given by $p$ when an internal belief is described by $q$. This is consistent with the KL divergence's role as the amount of information required to project (in the geometric sense) $p$ onto $q$. Though in this case, we usually think of $q$ as being an element of a submanifold in distribution space onto which we are projecting the point $p$. See Nielson's expository article \cite{N18} for additional details on projection.

The KL divergence is locally consistent with the Fisher metric \cite{A10,A16,N20} with the approximation,
\begin{equation*}
D_{KL}\left(p(\mathbf{x}\vert\bm{\eta})\vert p(\mathbf{x}\vert\bm{\eta}_0)\right) =\frac{1}{2} \mathbf{g}_{jk}(\bm{\eta}_0)\Delta\eta^j \Delta\eta^k + O(\norm{\bm{\Delta\eta}}^3),
\label{eqn:fisherkl}
\end{equation*}
holding for small distances. Thus, both the KL divergence and the Fisher metric measure square distance in nats (bits) and provide measures of ``surprise'' or information use in a Bayesian brain. In this case, the kinetic energy action measures incremental ``surprise'' from changes in belief, while the KL divergence measures ``surprise'' caused by the arrival of external information (beliefs). It is worth noting that there are several connections between information geometry, statistical mechanics, nonlinear dynamics and the Bayesian brain hypothesis \cite{F09,F10,KF11,LB13,KR15,BKMS17,BKR18,RBF18,CZCC21,AMTB22,FDSH23}.

\subsection{Dynamic Belief Model}
Using the foregoing observations, Goehle and Griffin \cite{GG24} integrated techniques from classical mechanics \cite{LL76} and information geometry to define an information theoretic mass-spring model \cite{GG24}. In this formulation,  the square-distance from Hooke's law that appears in the classical spring Lagrangian is replaced with the KL divergence, while the kinetic energy term is simply the kinetic energy Lagrangian. In the general case, the resulting information theoretic mass-spring Lagrangian is,
\begin{equation*}
    \mathcal{L} = \frac{m}{2}\mathbf{g}_{jk}(\bm{\eta}) \dot{\eta}^j\dot{\eta}^k - \frac{k}{2}D_{KL}\left(\bm{\eta}'\vert \bm{\eta}\right),
\end{equation*}
where $\bm{\eta}'$ is the fixed (or varying) origin of the information theoretic mass-spring system. Here $k$ and $m$ are spring and mass constants, respectively.

We can interpret this through the lens of Landauer's principle. The corresponding Hamiltonian is given by the Legendre transform as,
\begin{equation*}
    \mathcal{H} = \frac{1}{2m}g^{ij}\xi_i\xi_j + \frac{k}{2}D_{KL}(\bm{\eta}'|\bm{\eta}),
\end{equation*}
where $\bm{\xi}$ is the vector of conjugate momenta. As expected, potential ``energy'' is given by the Kullback-Liebler divergence, which gives (a kind of) projection (square) distance \cite{N18} from  $\bm{\eta}'$ to $\bm{\eta}$ just as the square Euclidean distance is used in Hooke's law. The kinetic ``energy'' is (precisely) the kinetic energy defined by the Fisher metric but expressed in terms of the conjugate momenta. As the system oscillates, we can think of (real) energy (in joules) being expended in processing the information needed to change an internal belief (kinetic energy) and then (real) energy being expended in interpreting external information parameterized by $\bm{\eta}'$ in terms of the moving internal belief parameterized by $\bm{\eta}$. In this case, information is conserved but real energy is expended. In \cite{GG24}, Griffin and Goehle hypothesize that nature wisely incorporates a damper (or friction) to prevent continued oscillations. We show in the sequel that an observed price can serve this same purpose.

For the Bernoulli distribution with parameter $q$, the kinetic energy term is particularly simple with,
\begin{equation*}
    T(q) = \frac{1}{2}\frac{1}{q(1-q)}\dot{q}^2.
\end{equation*}
Likewise, if $q'$ is another Bernoulli parameter, the Kullback-Leibler divergence gives the potential function,
\begin{equation*}
    V(q) = D_{KL}(q', q) = q' \log\left(\frac{q'}{q}\right) + (1-q')\log\left(\frac{1-q'}{1-q}\right).
\end{equation*}
We use these functional forms to model agent belief in the remainder of this paper, since any agent's belief about a price in a binary option market is just a Bernoulli variable describing the outcome of a binary event \cite{M06,WZ06}. Interestingly, a damped version of this model recovers the classic Friedkin-Johnson opinion dynamics model \cite{FJ90, SGSR18,G21,GSJ22} as a second order approximation \cite{GG24}.

\section{Pure Market Model}\label{sec:TheoreticalResults}
Assume there are $N$ agents in the market, and that these agents only observe the spot price $p$. We may assume that $\rho_i$ (Agent $i$'s belief about the price) is a (time-varying) Bernoulli parameter. Using a mass-spring model \cite{GG24}, the Lagrangian for agent $i$ is,
\begin{equation*}
    \mathcal{L}_i = \frac{m_i}{2\rho_i(1-\rho_i)}\dot\rho_i^2-\frac{k_i}{2}D_{KL}(p,\rho_i),
\end{equation*}
where $p$ is the price of Asset 1. Here, each agent's belief is drawn toward the price but also affected by its kinetic energy, which incorporates Fink's oscillating belief model \cite{KFWV96,FKHM02,CF08}. Assuming the agents are independent, the complete system Lagrangian is then, 
\begin{equation*}
    \mathcal{L} = \sum_i \mathcal{L}_i = \sum_i \left[\frac{m_i}{2\rho_i(1-\rho_i)}\dot\rho_i^2 - \frac{k_i}{2}D_{KL}(p, \rho_i)\right].
\end{equation*}
The system Hamiltonian follows from the Legendre transform in the usual way \cite{LL76},
\begin{equation}
    \mathcal{H}=\underbrace{\sum_i{\frac{\rho_i(1-\rho_i)}{2m_i}{\gamma_i^2}}}_{\text{Total Kinetic Energy}}+\underbrace{\sum_i\frac{k_i}{2}D_{KL}(p,\rho_i)}_{\text{Total Potential Energy}},
    \label{eqn:Hamiltonian1}
\end{equation}
where,
\begin{equation*}
    \gamma_i = \frac{\partial \mathcal{L}}{\partial \dot{\mathcal{\rho}}_i} = \frac{\partial \mathcal{L}_i}{\partial \dot{\mathcal{\rho}}_i} = \frac{m_i}{\rho_i(1-\rho_i)}\dot{\rho}_i,
\end{equation*}
is the conjugate (information or belief) momentum of $\rho_i$. As expected, the Hamiltonian can be separated into its kinetic and potential energy components. As before kinetic energy and potential energy as represented are information quantities whose total is conserved, while ``real'' energy is expended by the agent in processing input information from the observed price and the alteration of its beliefs. Notice as $\rho_i$ approaches the belief boundary, i.e., $0$ or $1$, both momentum and kinetic energy approach infinity. This is a function of the hyperbolic nature of the underlying Riemannian manifold and imposes a natural limit on belief; i.e., agents can never be perfectly certain of an outcome.

\begin{figure}[htbp]
\centering
\includegraphics[width=0.45\columnwidth]{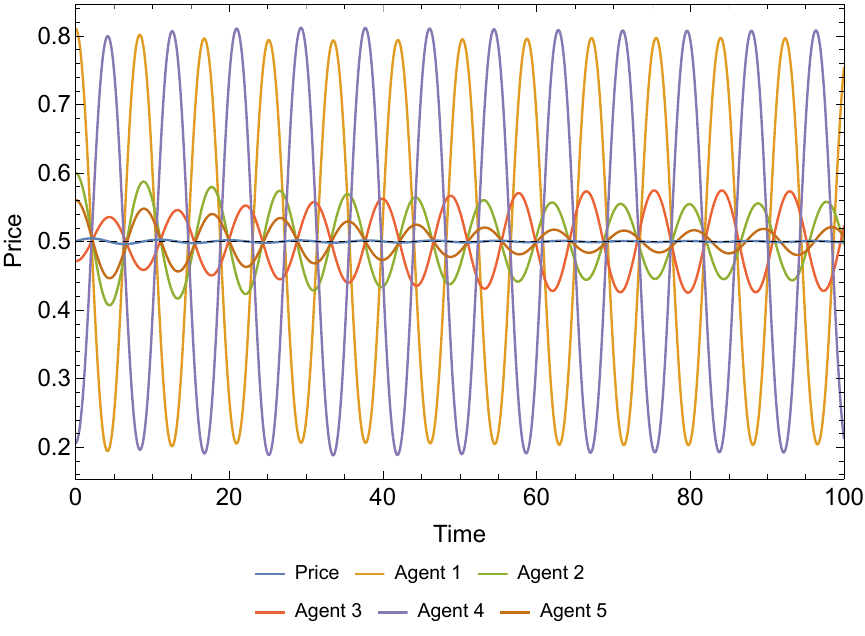} \quad
\includegraphics[width=0.45\columnwidth]{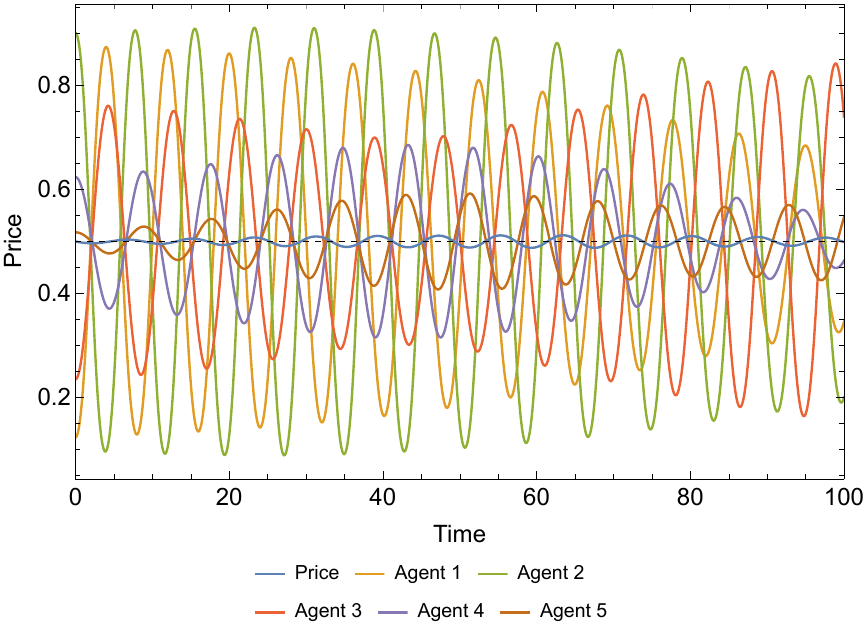}
\caption{(Left) A small market simulation consisting of five identical agents with random starting beliefs and no momentum ($\gamma_i(0) = 0$ for all $i$). The dashed line represents the starting price $p_0 = 0.5$ and the blue line is the oscillating price. (Right) A second example with the same initial momenta and starting price, but different initial conditions, showing the impact of initial conditions on agent belief and spot-price. For both plots, $m_i = k_i = Q_i = 1$.}
\label{fig:1}
\end{figure}

Complete system dynamics can then be derived from the Hamiltonian and the price dynamics given in \cref{eqn:pdot}. This yields a $2N+1$ dimensional system of differential equations,
\begin{equation}
\left\{
\begin{aligned}
    &\dot{\rho_i}=\frac{\partial\mathcal{H}}{\partial\gamma_i}=\frac{\rho_i(1-\rho_i)}{m_i}\gamma_i\\
    &\dot{\gamma_i}=-\frac{\partial\mathcal{H}}{\partial\rho_i}=\frac{k_i(p-\rho_i)}{2\rho_i(1-\rho_i)}-\frac{(2\rho_i-1)\gamma_i^2}{m_i}\\
    &\dot{p} = \beta p(1-p)\sum_i Q_i(\rho_i - p).
\end{aligned}
\right.
\label{eqn:DynamicsA}
\end{equation}
Notice this is an idealised market, with agents receiving signals through the spot price only. Example dynamics for a small market with five agents are shown in \cref{fig:1}. In this figure, all agents start with zero momentum, i.e., $\gamma_i(0) = 0$ for all $i$. The left and right figure components show the effect of different (random) starting conditions for $\rho_i(0)$. Price is shown in blue with the dashed line showing the initial price and oscillation around it. For both market realisations, $m_i = k_i = Q_i = 1$.

The dynamics in \cref{eqn:DynamicsA} have an infinite set of fixed points $\rho_i^* = p^*$ and $\gamma_i^* = 0$ for some arbitrary price $p^* \in [0,1]$ and these are the only fixed points. To see this, consider the right-hand-side of $\dot{\rho}_i$. Necessarily $\rho_i \in (0,1)$, so $\dot{\rho}_i = 0$ if and only if $\gamma_i = 0$. The resulting fixed point equations for $\dot{\gamma}_i$ have form,
\begin{equation*}
    \frac{\partial\mathcal{H}}{\partial\rho_i}=\frac{k_i(p-\rho_i)}{2\rho_i(1-\rho_i)} = 0,
\end{equation*}
which has unique solution $\rho_i = p$. This immediately makes $\dot{p} = 0$, with $p$ a free variable, thus leading to the infinite set of fixed points of the given form.

The behaviour of the market varies based on the number of traders and their relative equalities. Traders are identical if they have equivalent $m$, $k$, and $Q$ values. In the degenerate market with a single trader, the Jacobian matrix at any fixed point $\rho_1 = p$, $\gamma_1 = 0$ is given by,
\begin{equation*}
    \mathbf{J} = \begin{bmatrix}
 0 & \frac{(1-p) p}{m_1} & 0 \\
 -\frac{1}{2} k_1 \left(\frac{1}{p}+\frac{1}{1-p}\right) & 0 & \frac{1}{2} k_1
   \left(\frac{1}{p}+\frac{1}{1-p}\right) \\
 \beta  (1-p) p & 0 & -\beta  (1-p) p
\end{bmatrix}.
\end{equation*}
This matrix has eigenvalues,
\begin{align*}
    &\lambda_1 = 0\\
    &\lambda_{2,3} = \frac{-\beta  m_1 p(1-p)
   \pm\sqrt{m_1 \left(\beta ^2 m_1 (p-1)^2 p^2-2 k_1\right)}}{2 m_1}.
\end{align*}
This is an attracting hyperbolic fixed point with the slow manifold ($\lambda_1 = 0$) corresponding to movement along the infinite line of fixed points. Thus, the price acts as a natural damper in the market, and there are an infinite number of asymptotically stable fixed points. In systems with more than one trader, the price also acts as a damper on the system, but interestingly, the presence of identical traders will create a non-trivial centre manifold leading to oscillations.

\begin{figure}[htbp]
\centering
\includegraphics[width=0.45\textwidth]{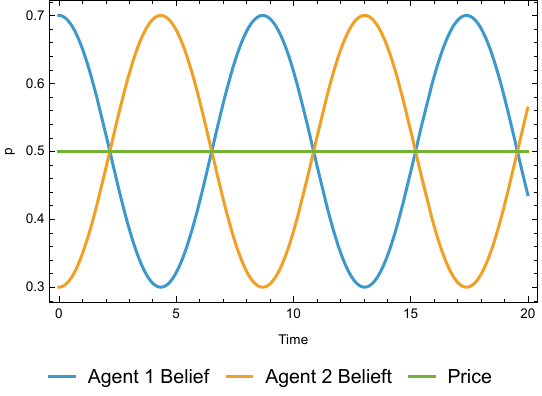}
\caption{An example of the centre manifold that can occur in these dynamics using a two agent market.}
\label{fig:Centre}
\end{figure}

For a system with $N > 1$ traders, the Jacobian matrix at a fixed point $\rho_i = p$ and $\gamma_i = 0$ for a $i \in \{1,\dots,N\}$ has characteristic polynomial in terms of $\lambda$,
\begin{equation*}
    -\frac{\lambda}{2^Nm^N}(k+2m\lambda^2)^{N-1}(k+2Nmp(1-p)\beta\lambda+2m\lambda^2).
\end{equation*}
The roots are,
\begin{align*}
    &\lambda_1 = 0\\
    &\lambda_{2,3} = \pm i\sqrt{\frac{k}{2m}}\\
    &\lambda_{4,5} = -\frac{5}{2} \beta  p(1-p) \pm\frac{\sqrt{m \left(25 \beta ^2 m (p-1)^2 p^2-2 k\right)}}{2m}.
\end{align*}
Eigenvalues $\lambda_2$ and $\lambda_3$ have multiplicity $N-1$ each, suggesting a $2N-2$ dimensional centre manifold, while the real parts of $\lambda_{4,5}$ are negative, indicating a two-dimensional stable manifold. While not constituting a formal proof, if $p$ is fixed (i.e., on a nullcline $\dot{p} = 0$), then the dynamical system becomes purely Hamiltonian and Liouville's theorem can be invoked. We illustrate this in \cref{fig:Centre}. In the case of two identical traders, when $\gamma_1(0) = \gamma_2(0)$ and $\rho_1(0) = \frac{1}{2}+\epsilon$ and $\rho_2(0) = \frac{1}{2}- \epsilon$, it is straightforward to see that, 
$\dot{p} = \dot{\rho}_1 = \dot{\rho}_2$ and $\dot{\gamma}_1 + \dot{\gamma}_2 = 0$. Thus, the total conjugate momentum is conserved, the system remains on a nullcline $\dot{p} = 0$ and the Hamiltonian $\mathcal{H}$ becomes a conserved quantity and the resulting system clearly exhibits a centre manifold as shown in \cref{fig:Centre}. 
The interaction of the slow manifold (the line of fixed points corresponding to $\lambda_1 = 0$), the stable manifold, and the centre manifold can create surprising stability and instability with the price frequently tending to $p^* = \tfrac{1}{2}$, even in systems with a large number of agents.

To test that assertion experimentally, we constructed a market with fifty identical agents with $m_i = k_i = Q_i=1$. In each numerical solution, lower $(l)$ and upper ($u$) bounds for the initial beliefs were fixed, and initial beliefs were chosen randomly between these bounds. That is, we chose 50 random values in $[l,u]$ and assigned each $\rho_i(0)$ to one of those values. The initial momenta were set to zero; i.e., $\gamma_i(0) = 0$ for all $i$. We tested under two conditions. In one condition, the initial price was set at $p_0 = \tfrac{1}{2}$ and in the second condition the initial price was set at $p_0 = \tfrac{3}{4}$. In all experiments, we set $\beta = 1$ and the dynamics evolved for 2000 time units.

\begin{figure}[htbp]
\centering
\includegraphics[width=0.95\textwidth]{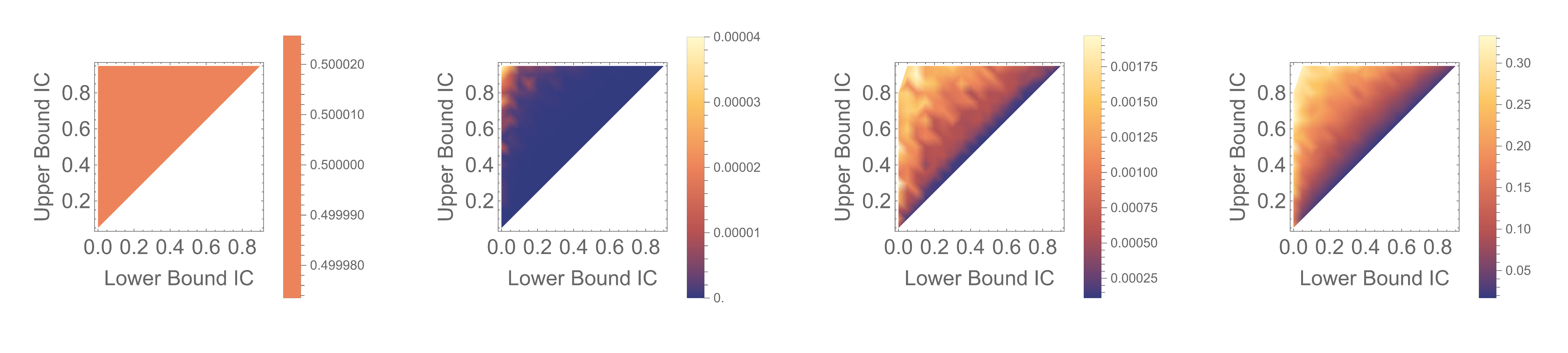}\\
\includegraphics[width=0.95\textwidth]{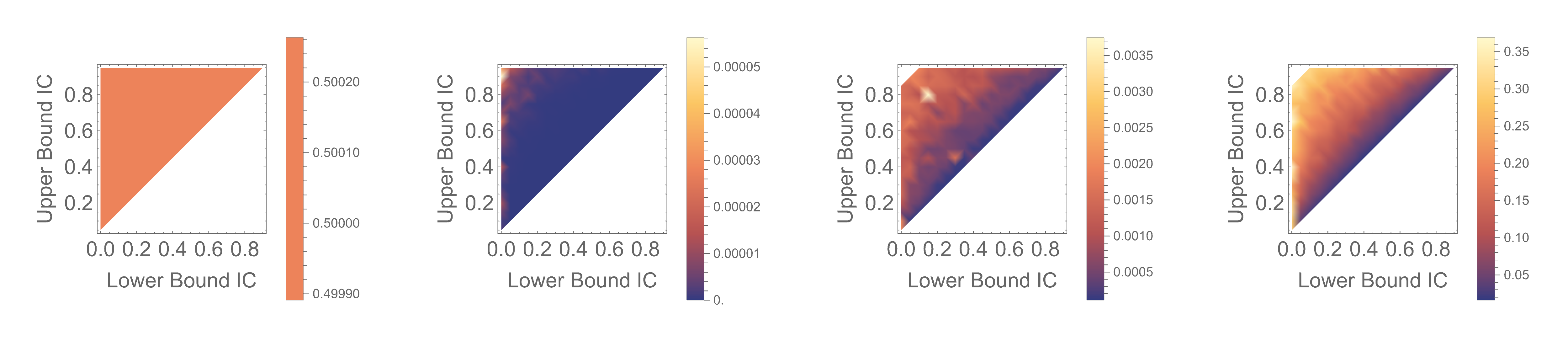}
\caption{(Top Row) For $p_0 = \tfrac{1}{2}$,  the density plot for $\mean{p}$ as a function $[l,u]$ shows convergence to $\tfrac{1}{2}$. Likewise, the density plot for $\mu$ shows this value is close $0$, suggesting convergence of the mean belief. The density plot for $\mean{p - \mean{p}}$ shows this value is close to zero for all starting conditions, suggesting the price converges. However, the density plot for $v$ is non-zero, suggesting the dynamics converge to the centre manifold and oscillate around their mean belief. (Bottom row) Similar results are shown for the starting price  $p_0 = \tfrac{3}{4}$.}
\label{fig:BasinOfAttraction}
\end{figure}

We computed the mean and variance for each trajectory during the last 500 time units as,
\begin{align*}
    &\mean{p}=\frac{1}{500}\int_{1500}^{2000} p(t)\,dt, \quad \mean{\rho_i} = \frac{1}{500}\int_{1500}^{2000} \rho_i(t)\,dt,\\
    &\mean{p-\mean{p}} = \frac{1}{500}\int_{1500}^{2000} p(t) - \mean{p}\,dt, \quad 
    \mean{\rho_i-\mean{\rho_i}} = \frac{1}{500}\int_{1500}^{2000} \rho(t) - \mean{\rho_i}\,dt.
\end{align*}
To summarise the results for the beliefs of all agents, let $\rho^*=\tfrac{1}{2}$. Define,
\begin{equation*}
    \mu = \max_i \left\lvert \mean{\rho_i} - \rho^* \right\rvert.
\end{equation*}
This is the maximum absolute difference in mean agent belief from $\rho^*$. If $\mu$ is close to zero, then mean agent belief converges to $\rho^*$ (as we hypothesize). Conveniently, this gives one value of $\mu$ for each $[l,u]$ interval. Likewise, define,
\begin{equation*}
    v = \max_i \mean{\rho_i-\mean{\rho_i}}.
\end{equation*}
This is the maximum trajectory variance over all agents. If $v$ is large and $\mu$ is close to zero, then this implies the agent beliefs must oscillate around the mean. If $\mean{p}$ is near $\tfrac{1}{2}$ and $\mean{p - \mean{p}}$ is small in absolute value, this implies the price convergenes (a we hypothesize). Density plots of the results as a function $[l,u]$ are shown in \cref{fig:BasinOfAttraction} for both test conditions with $p_0 = \tfrac{1}{2}$ shown in \cref{fig:BasinOfAttraction} (top) and $p_0 = \tfrac{3}{4}$ shown in \cref{fig:BasinOfAttraction} (bottom).
For both $p_0 = \tfrac{1}{2}$ and $p_0 = \tfrac{3}{4}$, the scales for $\mean{p}$ and $\mu$ are very narrow around $\tfrac{1}{2}$ and $0$ respectively, showing that both the price and the mean belief converge to $\tfrac{1}{2}$. The variance of the price is also close to zero, showing that price converges (as expected), while $v$ has a wider scale suggesting that a subset of the agents are converging to the centre manifold and oscillating around their mean value. 

\begin{figure}[htbp]
\centering
\includegraphics[width=0.45\textwidth]{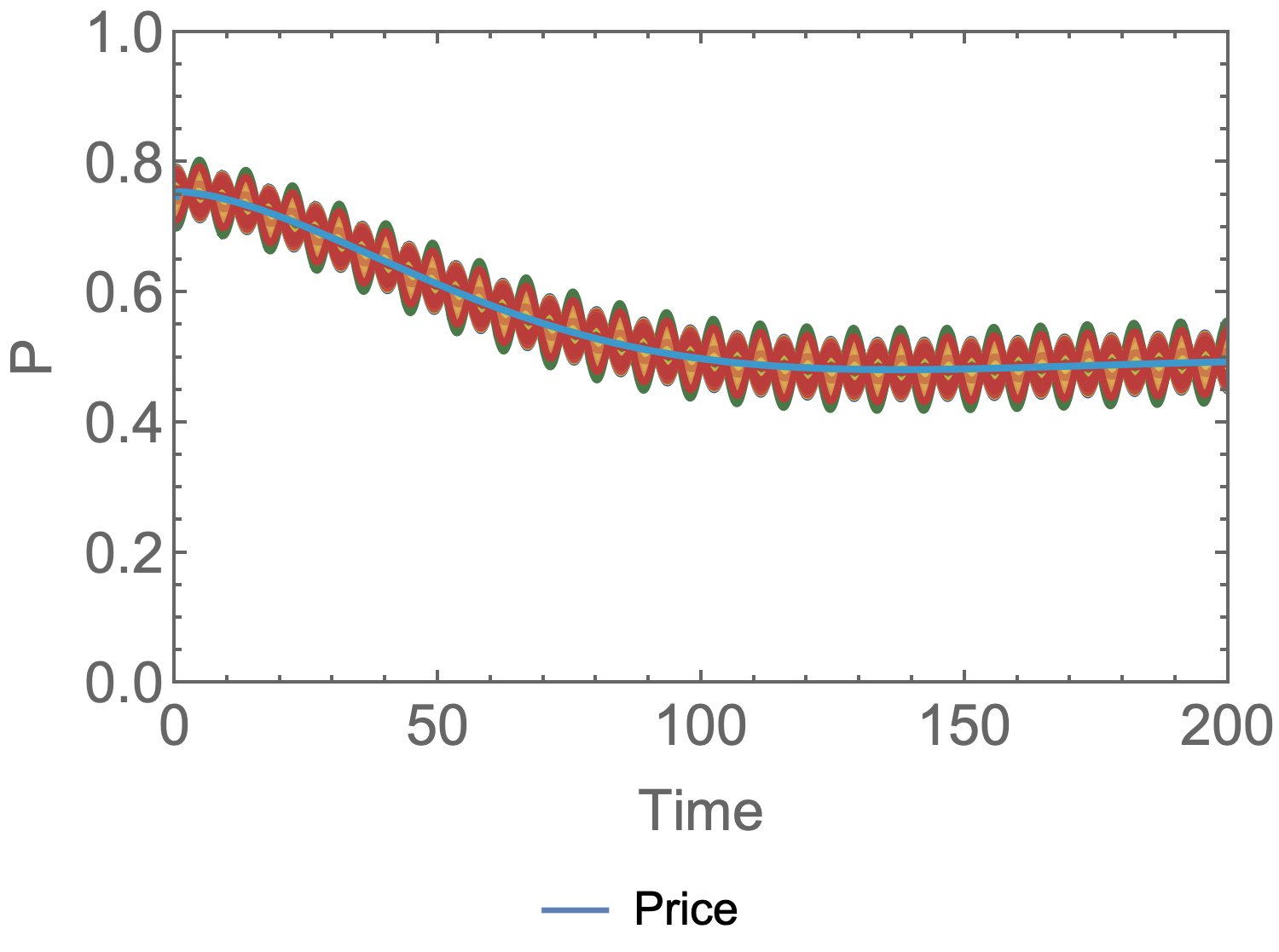} \quad
\includegraphics[width=0.45\textwidth]{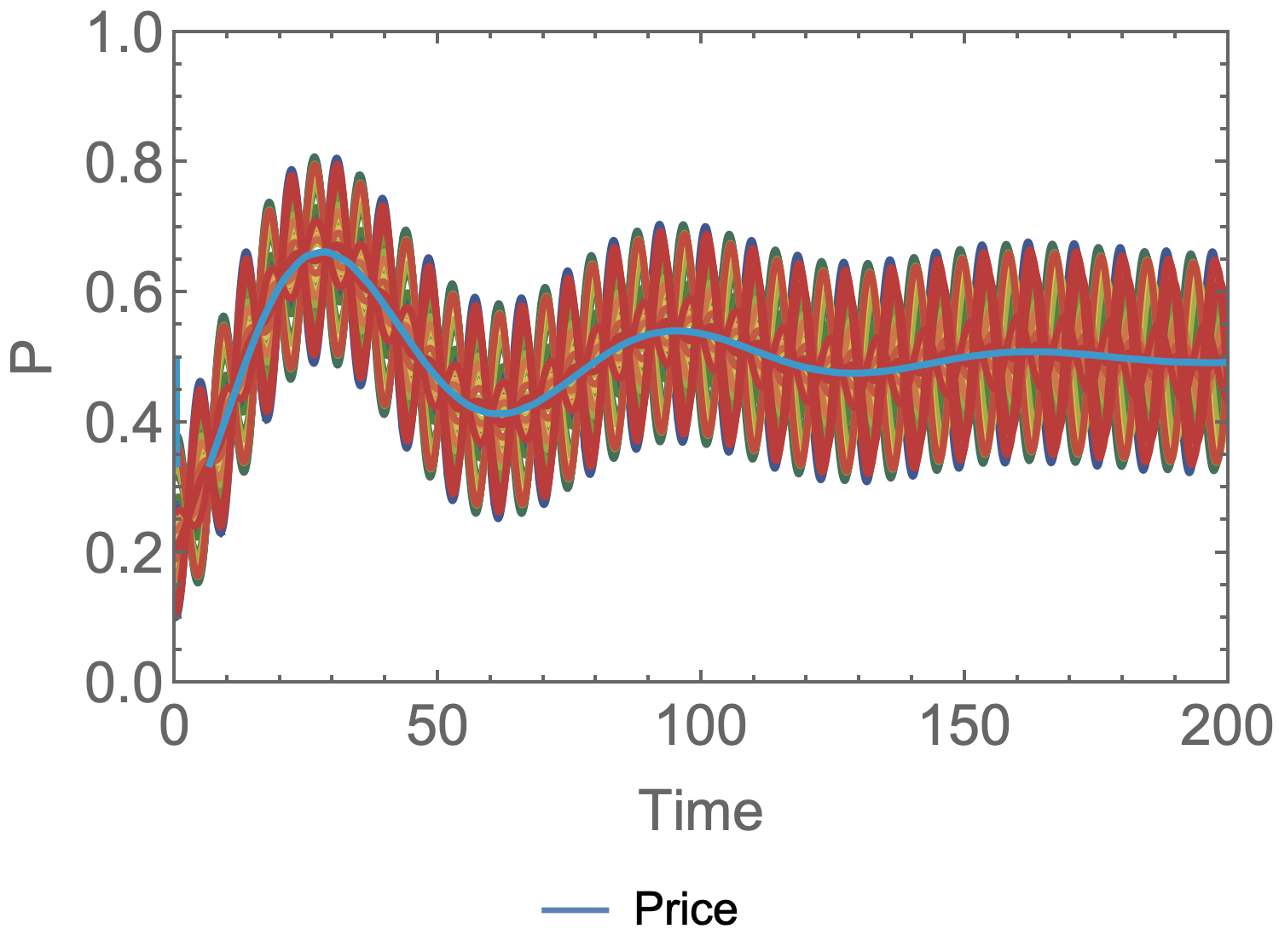}
\caption{(Left) Illustration of the geometry pushing the price toward $p^* = \tfrac{1}{2}$. (Right) Inconsistent price and initial conditions, converging to the basin of attraction around $p^*$.}
\label{fig:50Agents}
\end{figure}

Interestingly, there is very little impact on the output as a result of the price starting position, suggesting a global basin of attraction around the fixed point $p^* = \rho_i^* = \tfrac{1}{2}$ and $\gamma_i^* = 0$ for all $i$. This behaviour most likely arises as a result of the geometry of the Riemannian manifold on which the dynamics evolve, with regions of high curvature ($\rho_i \approx 0$ and $\rho_i \approx 1$) effectively pushing the dynamics toward the centre of the space. This is illustrated in \cref{fig:50Agents} (left). 
The basin of attraction is further illustrated in \cref{fig:50Agents} (right) where the price and agent belief are begun in inconsistent regions and converge to a fixed $p^*=\tfrac{1}{2}$.

\begin{figure}[htbp]
\centering
\includegraphics[width=0.45\textwidth]{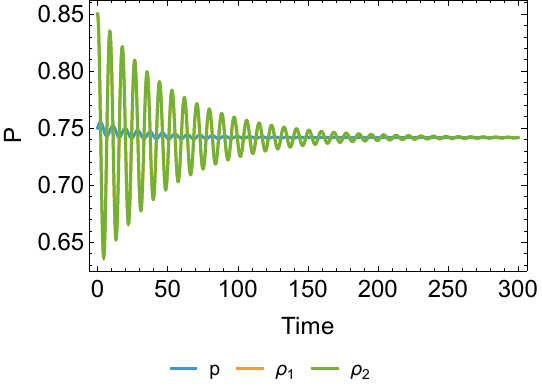}
\caption{An alternate basin of attraction for two agents.}
\label{fig:OtherBasin}
\end{figure}

It is worth noting that this global attraction is very much a function of the number of traders (system dimension) and initial conditions. For example, in the two agent system when we initialise $\rho_1(0) = \rho_2(0) = 0.85$, $\gamma_1(0) = \gamma_2(0) = 0$ and $p_0 = 0.75$, with $\beta = \tfrac{1}{10}$, the price converges above $\tfrac{1}{2}$ as shown in \cref{fig:OtherBasin}.

\subsection{Unequal Traders}
We now introduce a simple form of inequality into the model and study the resulting dynamics. Heretofore, we have assume all traders were identical. Without loss of generality, we now suppose that trader 1 differs from all other (identical) traders. That is, we assume that $m_i = m$ and $k_i = k$ for $i = 2,\dots,N$ and $m_1 \neq m$ and $k_1 \neq k$. For simplicity we still assume $Q_i = 1$ for all $i$. That is all traders have equal market power.
%Consider now the scenario with an unequal trader so that $m_i = m$ and $k_i = k$ for $i = 2,\dots,N$. 
We will show that this inequality has the potential to change the composition of the stable and centre manifolds. 

\begin{figure}[htbp]
\centering
\includegraphics[width=0.45\textwidth]{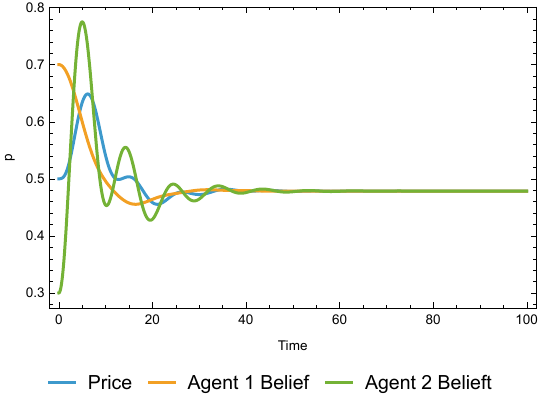} \quad 
\includegraphics[width=0.45\textwidth]{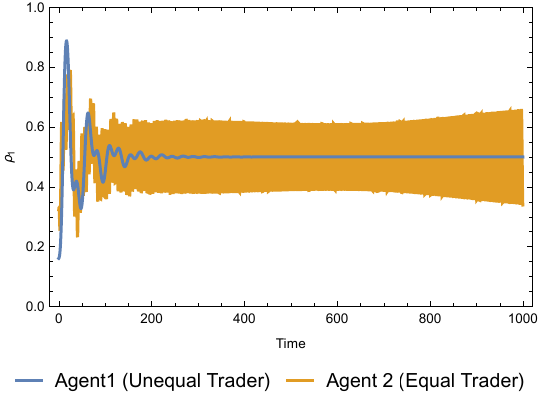}
\caption{(Left) Both price and beliefs convergence in a market with two unequal traders, as expected from theory. (Right) In a market with many identical traders and one atypical trader, the atypical trader converges asymptotically to a fixed belief, while the remainder of the traders converge to the centre manifold.}
\label{fig:StableManifold}
\end{figure}

The characteristic polynomial of the Jacobian matrix at the fixed point is given by,
\begin{equation*}
    -\frac{1}{2^Nm^{N-1}m_1}\lambda(k+2m\lambda^2)Q(\lambda),
\end{equation*}
where,
\begin{multline*}
    Q(\lambda) = kk_1 
    +[2(N-1)p{\beta}k_1-2(N-1)mp^2{\beta}k_1+2kp{\beta}m_1-2kp^2{\beta}m_1]\lambda
    +\\2mk_12km_1 \lambda^2+ 
    (4nmp{\beta}m_1-4mp^2{\beta}m_1)\lambda^3 + 4mm_1\lambda^4.
\end{multline*}
We can scale this quartic to have form,
\begin{equation*}
    \tilde{Q}(\lambda) = \lambda^4 + a_1\lambda^3+a_2\lambda^2+a_3\lambda+a_4,
\end{equation*}
and applying a combination of the Routh-Hurwitz criteria and the sufficient conditions for a pair of pure imaginary solutions to a quartic \cite{spiegel2009schaum, gopal2008control}. We test whether,
\begin{equation*}
    a_3^2+a_1^2a_4=a_1a_2a_3.
\end{equation*}
% We'll test the nature of the quartics roots by testing:
% \begin{equation*}
%     a_3^2+a_1^2a_4=a_1a_2a_3
% \end{equation*}
%
% The next case we will analyze is if there exists one trader with a different weight, $m_1$ and $k_1$, than all other traders. This "oddball" trader will add inequality to the market, and potentially alter the behavior of the system. The quartic for this case when there are N traders is:
% \begin{equation*}
%     a_0\lambda^4+a_1\lambda^3+a_2\lambda^2+a_3\lambda+ca
% \end{equation*}
% where,
% \\
% $
%      a_0=\frac{-4mm_1}{-4mm_1}=1\\
%      a_1=\frac{-4nmp{\beta}m_1+4mp^2{\beta}m_1}{-4mm_1}\\
%      a_2=\frac{-2mk_12km_1}{-4mm_1}\\
%      a_3=\frac{-2(n-1)p{\beta}k_1+2(n-1)mp^2{\beta}k_1-2kp{\beta}m_1+2kp^2{\beta}m_1}{-4mm_1}\\
%      a_4=\frac{-kk_1}{-4mm_1}
%  $
%  \\
% Using the same method of analysis as the equivalent market, we're left with:
% \begin{multline*}
%     \frac{n(-1+p)^2p^2{\beta}^2(mk_1+km_1)(m(-1+n)k_1+km_1)}{4m^2{m_1}^2}=\\
%     \frac{(-1+p)^2p^2{\beta}^2(m^2(-1+n)^2{k_1}^2+km-2+n(2+n))k_1m_1k^2{m_1}^2}{4m^2m_1^2}
% \end{multline*}
When simplified this gives,
\begin{multline*}
   {N(mk_1+km_1)(m(N-1)k_1+km_1)}=\\
  {(m^2(N-1)^2{k_1}^2+km-2+n(2+N))k_1m_1k^2{m_1}^2},
\end{multline*}
which holds precisely when 
\begin{equation*}
    (N-1)(mk_1-km_1)^2=0,
\end{equation*}
or more specifically we require,
\begin{equation*}
    \frac{k_1}{k} = \frac{m_1}{m}.
\end{equation*}
That is, the dimension of the centre manifold (at least about the nullcline $\dot{p} = 0$) is preserved precisely if the change in the mass of a trader is offset by a proportionate change in the corresponding spring constant. Thus, if the asymmetry introduced by the unequal trader isn't too great, then the system will continue to have purely imaginary roots and a centre manifold of dimension $2N-2$. However, if the asymmetry introduced by the unequal trader is too great, the dynamics have a centre manifold reduced in dimension by 2. Surprisingly, the Routh-Horowitz criteria suggest that the asymmetry increases the dimension of the stable manifold. That is, the additional non-imaginary roots have negative real-part. Thus, for this simple market dynamic, inequality and trader diversity have the potential to increase the dimension of the stable manifold, increasing the likelihood that both the price and agent beliefs will converge. This is illustrated for two unequal traders in \cref{fig:StableManifold} (left) where we use the same dynamical system as in \cref{fig:Centre}, but change the mass of agent 1 to $m_1 =10$. In \cref{fig:Centre}, we saw belief oscillation. In comparison, in \cref{fig:StableManifold} (left) we see asymptotic convergence to the stable manifold in belief as expected from the theoretical results.
This phenomenon is present in larger markets with one anomalous trader. This is shown in \cref{fig:StableManifold} (right) where we use fifty agents. Here agent 1 has mass $m_1 =\tfrac{1}{2}$ and for $i \neq 1$ we have $m_i = m = 1$. Agent1's belief asymptotically converges to a fixed point while the other (equal) trader's beliefs seem to converge to the centre manifold and oscillate, as before in \cref{fig:1,fig:Centre}.

\subsection{Interpretation of Results}
The results in this section suggest an interesting conclusion: the geometry of the Riemannian manifold along with the price-based spring dynamics, create a semi-dissipative system in which price generally moves toward $p^* = \tfrac{1}{2}$ and converges to some fixed value that is dependent on initial conditions, though generally equal to $p^*$ for larger agent systems. Once $\dot{p} \approx 0$, agent dynamics are defined by a Hamiltonian system and the trajectories evolve along the system's centre manifold. Introduction of asymmetry or inequality into the market has the potential to increase the dimension of the stable manifold, though in general market asymmetries or inequalities among the agents do not dramatically change these observations. A formal proof of these results is left to future work, as it is outside the scope of this paper. 

While the dynamics are an interesting example of a Hamiltonian system coupled to a type of damper (the price dynamics acting as the damper), this is clearly neither how real traders nor markets function. In the next sections, we introduce exogenous information to the markets and study the effect that purchasing power inequality can have on the markets, illustrating the evolution of belief reinforcing market bubbles.

\section{External Information and Trader Communication}
\label{sec:ExternalInformation}
Real-world traders are both exposed to external information and interact with each other (outside of observing the price). We can modify the Hamiltonian in \cref{eqn:Hamiltonian1} as,
\begin{multline}
    \mathcal{H} =\sum_i\left({\frac{\rho_i(1-\rho_i)}{2m_i}{\gamma_i^2}}+\frac{k_i}{2}D_{KL}(p,\rho_i)\right)+\\
    {\sum_j{\frac{k_{ij}}{2}D_{KL}(\rho_j,\rho_i)}}+\frac{r_i}{2}\sum_{i}D_{KL}(\eta_i,\rho_{i}).
\label{eqn:Hamiltonian2}
\end{multline}
Here $\eta_i$ is an external signal received by agent $i$ and $k_{ij}$ is a (spring) weight assigned to agent $j$'s (private) belief about the price by agent $i$. Goehle and Griffin studied \cref{eqn:Hamiltonian2} without price dependence in \cite{GG24}, showing that when an appropriate dissipation term is included, the model can recover a variation of the Friedkin-Johnson opinion dynamics \cite{FJ90} with peer pressure \cite{SGSR18,GSJ22}. In the market context, external information could be those received by the trader from his/her corporation, a news source, or a prior belief that continues to exert force on the instantaneous belief \cite{GG24}.

\subsection{Communication between Agents}
\label{sec:AgentCommunication}
\begin{figure}[htbp]
\centering
\includegraphics[width=0.95\textwidth]{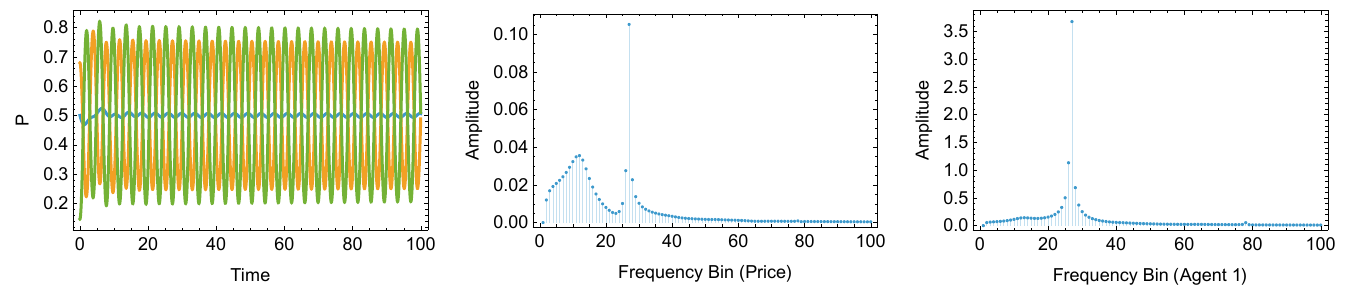}
\caption{(Left) The dynamics of two communicating agents. (Centre) The Fourier spectrum of the price. (Right) The Fourier spectrum of Agent 1. Both spectra show non-chaotic behaviour.}
\label{fig:TwoAgentsComm}
\end{figure}

\begin{figure}[htbp]
\centering
\includegraphics[width=0.95\textwidth]{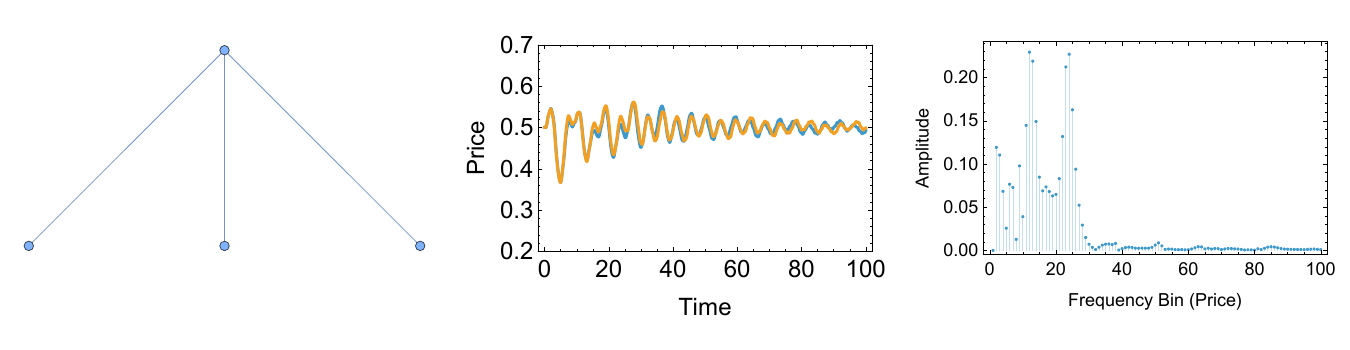}
\caption{(Left) A communications graph. (Centre) Sensitive dependence on initial conditions are illustrated. (Right) The Fourier spectrum of the price indicate complex, but not chaotic behaviour.}
\label{fig:FourMarket}
\end{figure}

\begin{figure}[htbp]
\centering
\includegraphics[width=0.95\textwidth]{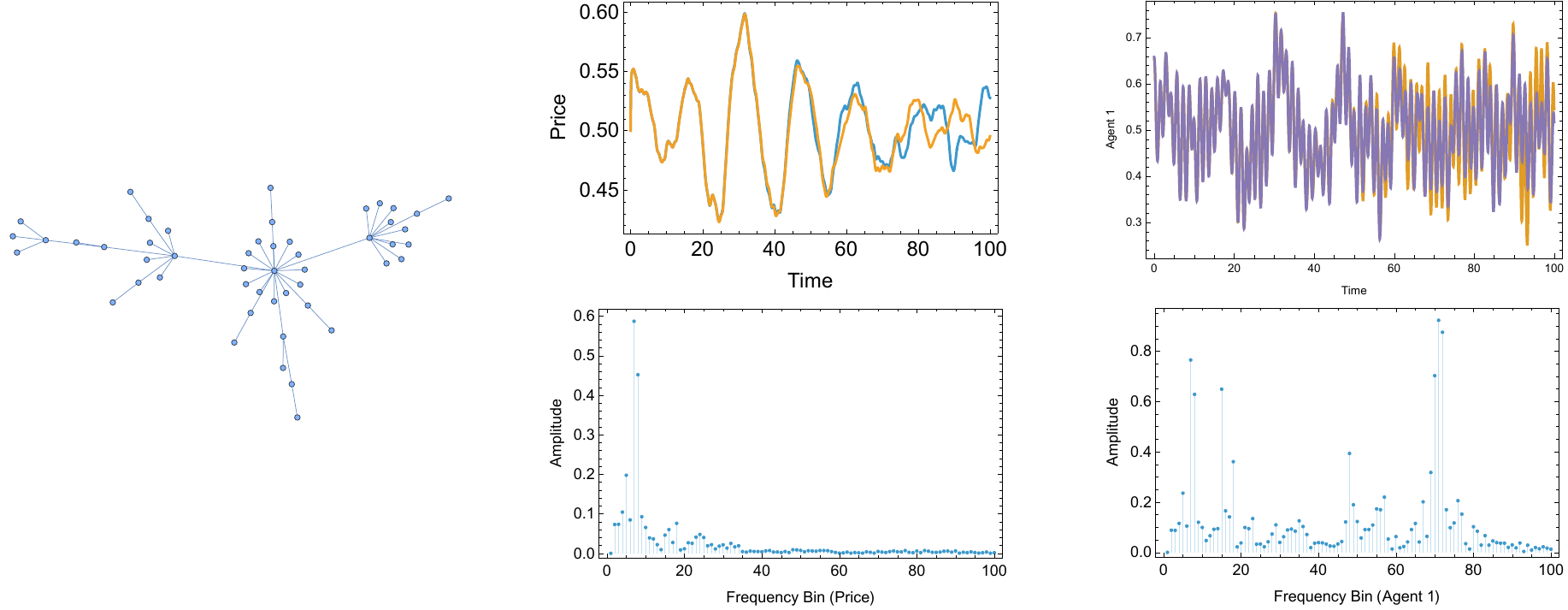}
\caption{(Left) The Barab{\'a}si-Albert communications graph. (Centre) Sensitive dependence on initial conditions are illustrated. (Right) The Fourier spectrum of the price indicate complex, but not chaotic behaviour (again)}
\label{fig:FiftyMarket}
\end{figure}

% From our Pure Market Model, we can add communication between agents through the inclusion of the term,$\sum_j{k_{ij}D_{KL}(\rho_j,\rho_i)}$, making our new Hamiltonian, $\mathcal{H}_1$:
% \begin{equation*}
%     \mathcal{H}_1=\sum_i\left({\frac{\rho_i(1-\rho_i)}{2m_i}{\gamma_i^2}}+D_{KL}(p,\rho_i)\right)+
%     {\sum_j{k_{ij}D_{KL}(\rho_j,\rho_i)}}
% \end{equation*}
Consider the case where $r_i = 0$ (i.e., there is no external information). Agents communicate according to some network whose weighted adjacency matrix is given by $k_{ij}$. The resulting dynamical system shares the same fixed points as the system with no communication, but the behaviour of the dynamics near  these fixed points is quite different, resulting in a price that does not converge. Such dynamics are illustrated in \cref{fig:TwoAgentsComm} (left) for two communicating agents with $k_{12} = k_{21} = 1$ and $k_i = m_i = 1$ for all agents.
Interestingly, Goehle and Griffin \cite{GG24} show that the Hamiltonian system built from \cref{eqn:Hamiltonian2} when $k_i = 0$ i.e., with no market model, exhibits weak chaos. In this case, the Fourier spectra shown in \cref{fig:TwoAgentsComm} (centre) and \cref{fig:TwoAgentsComm} (right) clearly show that these dynamics are not chaotic and have a fundamental frequency. The dependence on price smooths the (weak) chaotic behaviour observed in \cite{GG24}. This behaviour is also exhibited in larger markets, as shown in \cref{fig:FourMarket} and \cref{fig:FiftyMarket} for markets of size four and fifty.

As can be seen in \cref{fig:FourMarket}, when more agents are introduced with fewer edges between agents, in this case, each vertex (agent) has degree 1, the model trajectories become more complex. We used a simple graph (technically a Barab{\'a}si-Albert  graph \cite{BA99}) as the communication model. The graph is shown in \cref{fig:FourMarket} (left).
To illustrate sensitivity to initial conditions, we introduced a small perturbation of size $0.005$ to $\rho_1(0)$. \cref{fig:FourMarket} (centre) shows the resulting divergences of the two price trajectories. Despite this sensitivity, the Fourier spectrum shows a wide band of frequencies, but does not exhibit the usual behaviour common in a chaotic signal. The complexity of the dynamics is shown more clearly in a fifty agent market shown in \cref{fig:FiftyMarket}, using a more complex Barab{\'a}si-Albert graph.
As in the four agent case, the system exhibits sensitive dependence to initial conditions, yet the Fourier spectra again show indications of quasi-periodicity with multiple frequencies, but not the characteristically broad spectra usually associated with chaotic behaviour. We hypothesize that this is a result of the damping property of the common price observed by all agents.

\subsection{External Information and Power Inequalities}
\begin{figure}[htbp]
\centering
\includegraphics[width=0.45\columnwidth]{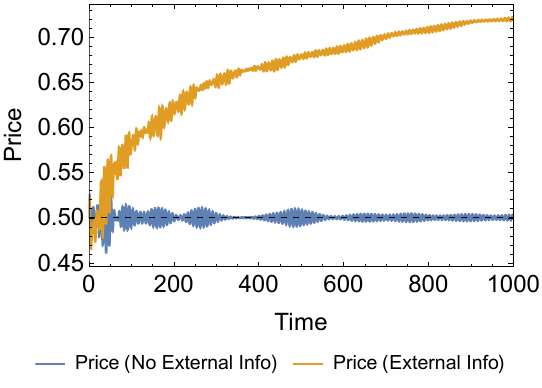}
\caption{This figure shows the price difference between two identical systems (10 traders, $P_0$=0.5) when one experiences external info and the other doesn't.}
\label{fig:5}
\end{figure}

\begin{figure}[htbp]
\centering
\includegraphics[width=0.45\columnwidth]{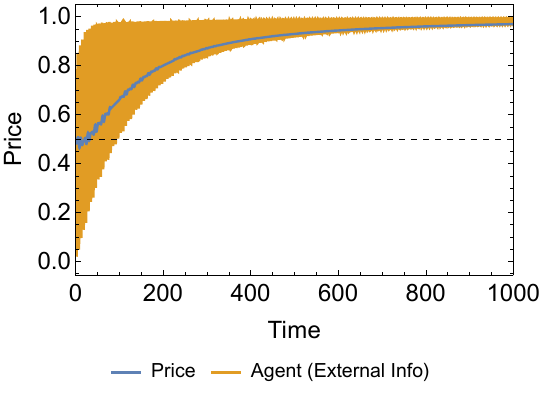} \quad \includegraphics[width=0.45\columnwidth]{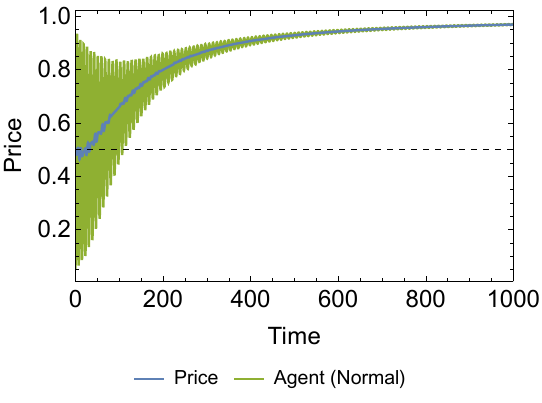}
\caption{Above's figures show the impact of an agent with external information (Left) on both price and another normal agent (Right) in a two agent model.}
\label{fig:6}
\end{figure}

In the real-world, a trader's (agent's) belief may be affected by external information sources or their own biases. This can affect both the spot price and other agent's beliefs, as illustrated in \cref{fig:5} in which 10 agents interact in a market and one agent (agent 1) is interacting with external information $\eta_1 = 1-e^{-.05t}$. This external information impacts all agents in the market (through the price changes caused by the externally influenced agent), and forces them to believe the asset is more valuable than it is regardless of price and communication information. This drives the price upward, creating a novel behaviour compared to the previous two cases. This upward drive of the price is apparent in \cref{fig:5}.
The impact that this has on the other traders' belief is shown in \cref{fig:6}. Here, the trader receiving external information is artificially driving the price up and therefore also the belief of another agent in a two-agent model.
Notice the price and belief still tend to converge, but they no longer converge at $p^*=\tfrac{1}{2}$ because the external information is both non-constant and has substantially changed the equilibrium points and consequently the structure of the stable and centre manifolds.

\begin{figure}[htbp]
\centering
\includegraphics[width=0.45\columnwidth]{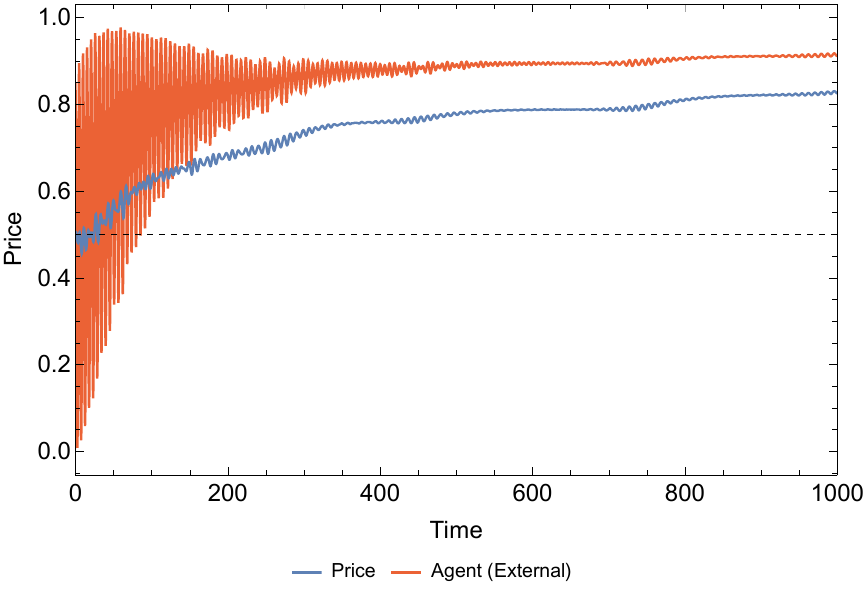} \quad \includegraphics[width=0.45\columnwidth]{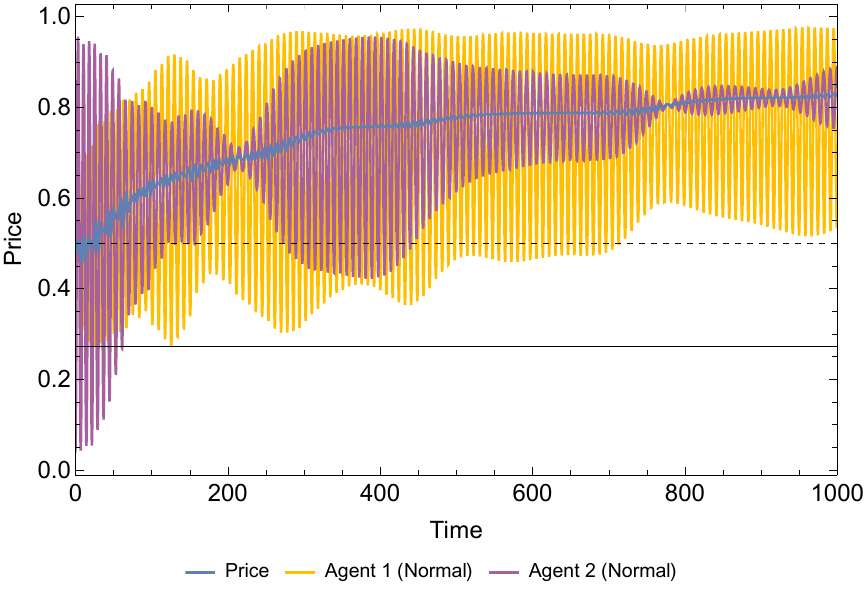}
\caption{Experiment highlights a model with 10 traders when one receives external information and has buying power $Q_{unequal}$=2 (Left) while the rest are normal with buying power $Q_{equal}$=1 (Right).}
\label{fig:7}
\end{figure}

When external information (or prior beliefs) are held by agents with exceptional purchasing power (i.e., markets with high trader inequality as given by high $Q_i$'s) the results can become disastrous, allowing us to model market bubbles and/or pump and dump schemes \cite{HC15}. We illustrate the impact of inequality in purchasing power with a ten agent market, in which one agent has double the purchasing power of all other agents. The agent with double the purchasing power has a time-varying signal,
\begin{equation*}
    \eta_1(t) = 1-\exp\left[-\tfrac{1}{20}t\right].
\end{equation*}
The resulting dynamics are illustrated in \cref{fig:7}.
Notice the belief of the agent with time-varying external information leads the market price, which increases as a result of that agents' purchasing. As expected, the means of the other agents' beliefs also track the price, causing agents to alter their beliefs as a result of interaction in the market. Effectively, external information (from any agent) is incorporated into the price, which then alters the belief of the other agents irrespective of whether they receive the information. This creates a feedback loop on the price, which could artificially raise the price above a reasonable level.

\begin{figure}[htbp]
\centering
\includegraphics[width=0.65\textwidth]{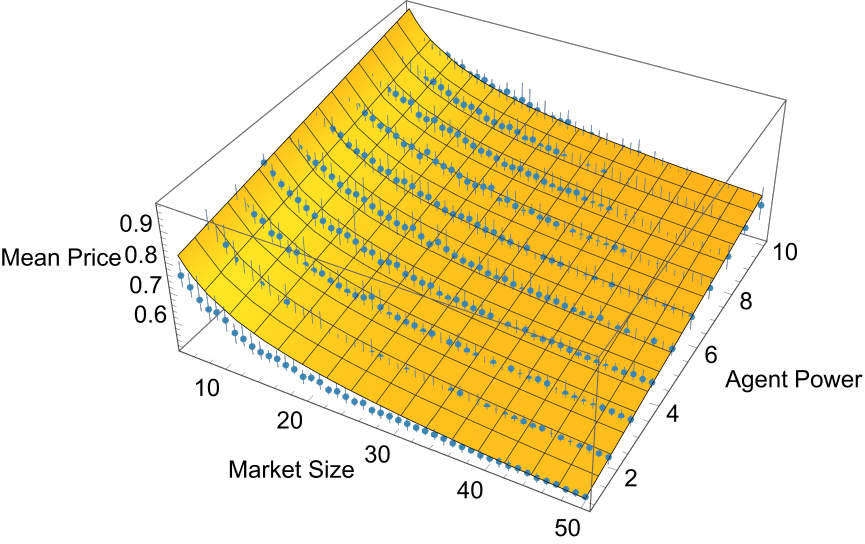}
\caption{Experimental results and fit showing the effect of a single agent with high purchasing power on a market of varying size.}
\label{fig:PowerExperiment}
\end{figure}

%%EXPERIMENTAL RESULTS
We constructed a simple experiment to measure the impact of a single powerful trader on markets of varying sizes. We varied $Q_1$, the purchasing power of agent 1 and set $\eta_1 = 0.9$. That is, the agent received a constant signal that the asset value was $0.9$ (a constant). There was no communication, so all agents received their information through the market, except for the agent with the external signal. Market size was varied from five to fifty agents and $Q_1$ was varied from one to ten. We ran the market to a final time $t_f = 500$ and computed,
\begin{equation*}
    \mean{p_f} = \frac{1}{100}\int_{400}^{500} p(t)\, dt.
\end{equation*}
Experimental results are illustrated in \cref{fig:PowerExperiment} along with the fit,
\begin{equation*}
    \mean{p_f} \sim \beta_0+ \frac{\beta_1}{N^{0.132}} + \beta_2Q,
\end{equation*}
with parameter table shown below.
\begin{equation*}
\begin{array}{l|llll}
 \text{} & \text{Estimate} & \text{Standard Error} & \text{t-Statistic} & \text{P-Value} \\
\hline
 1 & -0.33 & 0.005 & -73.64 & 0. \\
 \frac{1}{n^{0.132}} & 1.36 & 0.007 & 202.70 & 0. \\
 Q & 0.018 & 0.0001 & 139.59 & 0. \\
\end{array}
\end{equation*}
Surprisingly, within the parameter regime given, this model explained 93\% of the data variance ($r^2_\text{adj} = 0.93$). 
Because $\mean{p_f}$ must be between $0$ and $1$, this model must be an approximation (e.g. a partial Laurent series) for the true behaviour. While identifying a reasonable model for this is left to future research, it is surprising that the decay of the effect varies inversely with $N^{0.132}$ rather than $N$, suggesting that powerful agents might play an outsized role even in large markets. 

\begin{figure}[htbp]
\centering
\includegraphics[width=0.45\columnwidth]{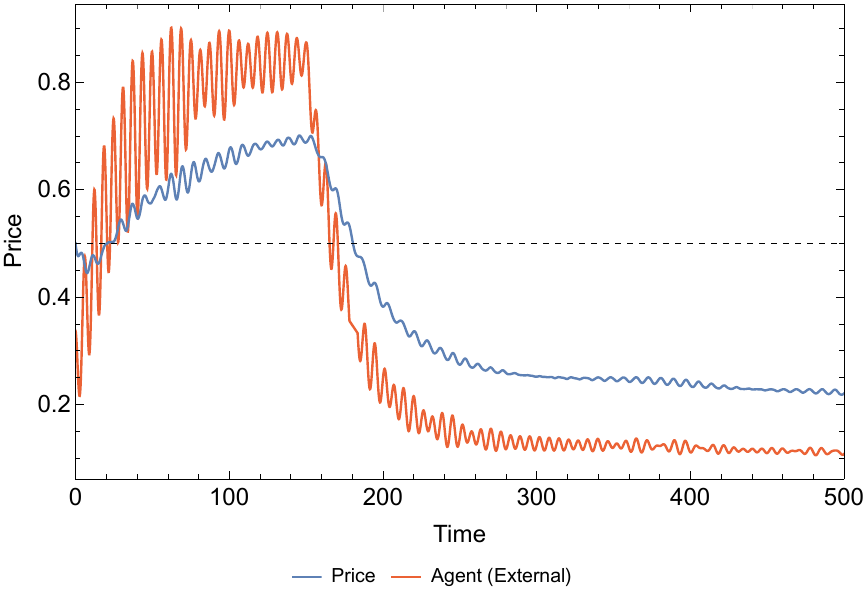} \quad \includegraphics[width=0.45\columnwidth]{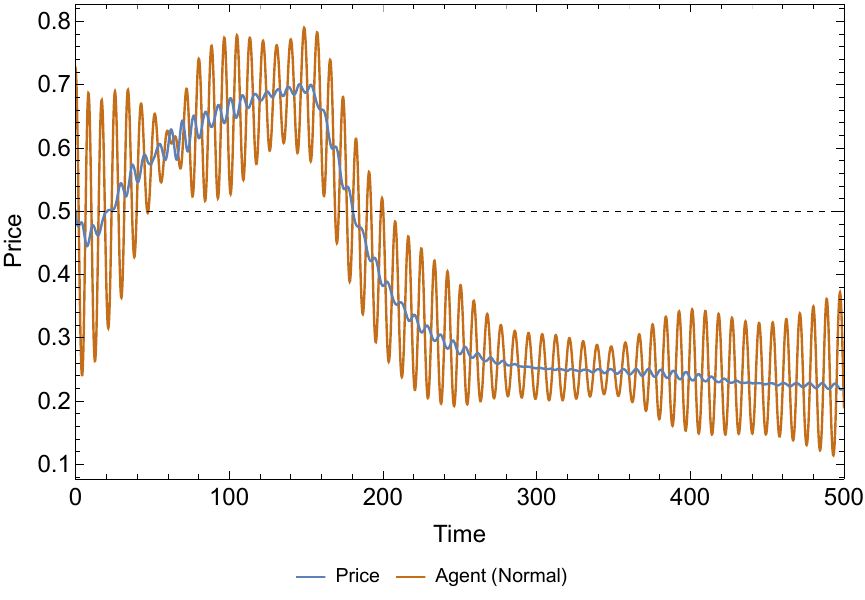}
\caption{Experiment above is a 10 trader market with an agent receiving external information and $Q_{unequal}$=2 (Left) and all other being normal with $Q_{equal}$=1 (Right).}
\label{fig:8}
\end{figure}

We can model a market bubble or a pump and dump scheme \cite{HC15} by considering the effect on the market when powerful agents suddenly alter their beliefs, either maliciously or as a result of Keynes' ``animal spirits''. We see the effect in \cref{fig:8} where the external signal is piecewise and switches from approaching 1 to approaching 0 after 150 time units; i.e.,
\begin{equation*}
    \eta_1 = \begin{cases}
        1 - \exp{\left[-\frac{t}{20}\right]} & \text{if $t < 150$}\\
        \exp{\left[-\frac{t}{20}\right]} & \text{if $t > 150$}.
    \end{cases}
\end{equation*}
These plots highlight the impact of having more purchasing power in a market. In certain cases, traders with higher purchasing power are capable of turning the entire market, even if their purchasing power is only twice that of their competitors. This ability to swing the market creates a feedback loop with the less powerful trader, in which the trader's belief is reinforced by their trades. This feedback loop can be seen in the volatility of the trader's beliefs over time. The powerful trader has significantly less volatility than the normal trader. The normal traders receive less feedback (only seeing the price) and therefore exhibit higher levels of volatility in their beliefs, while simultaneously tracking the more powerful trader.

%ADD IN THE SPECIAL CASE WITH EULER EQN WE DISCUSSED HERE******%
\section{Optimal Price Manipulation}\label{sec:Manipulation}
We can exploit the fact that the kinetic energy action of the Fisher metric gives the free entropy along a path of changing distribution (e.g., trader belief or price) to model a manipulative trader (or group of traders) who wish to move a market while causing the minimal amount of ``surprise'' (free entropy). This approach to controlling information was considered more generally by Griffin and Goehle in \cite{GG24a}. Consider the problem of moving a Bernoulli random variable along a free entropy geodesic. Heretofore, we have assumed the price dynamics to be given by \cref{eqn:pdot}. But in the case of intentional price manipulation, the price itself is the parameter and its Lagrangian is given by,
\begin{equation*}
\mathcal{L}_p = \frac{1}{2}\frac{\dot{p}^2}{p(1-p)}.
\end{equation*}
Griffin and Goehle \cite{GG23} show that a Bernoulli distribution follows geodesics given by, 
\begin{equation*}
   p^*(t) = \cos^2\left[\frac{1}{2} c_1 (t+c_2)\right],
\end{equation*}
where, 
\begin{equation*}
    c_1 =\frac{2 \left(\cos ^{-1}\left(\sqrt{p_f}\right)-\cos
   ^{-1}\left(\sqrt{p_0}\right)\right)}{T} \quad \text{and}  \quad c_2 =\frac{T \cos ^{-1}\left(\sqrt{p_0}\right)}{\cos ^{-1}\left(\sqrt{p_f}\right)-\cos
   ^{-1}\left(\sqrt{p_0}\right)}. 
\end{equation*}
Suppose that $\dot{\chi} = \dot{q}_1 - \dot{q}_0$, the term appearing on the right-hand side of the generic dynamics for $p$ in \cref{eqn:Rawqdot}. Then,
\begin{equation*}
    \dot{\chi} = \frac{1}{\beta}\frac{\dot{p}}{p(1-p)} = 
    -\frac{2 c_1 \csc \left(c_1 \left(c_2+t\right)\right)}{\beta }.
\end{equation*}
Integrating gives the open-loop asset purchase control law,
We can compute,
\begin{equation*}
    \chi = \frac{2 \tanh ^{-1}\left(\cos \left(c_1 \left(c_2+t\right)\right)\right)}{\beta }.
\end{equation*}
Without loss of generality, assume trader 1 is the manipulative agent using the purchasing strategy $\eta$ (to be defined in terms of $\chi$). Then the price dynamics are,
\begin{equation*}
\dot{p} = \beta p(1-p) \left[\underbrace{\eta + \sum_{i=2}^N Q_i(\rho_i - p)}_\chi\right].
\end{equation*}
Then,
\begin{equation*}
    \eta(p) = \chi - \sum_{i=2}^N Q_i(\rho_i - p) = \frac{2 \tanh ^{-1}\left(\cos \left(c_1 \left(c_2+t\right)\right)\right)}{\beta } - \sum_{i=2}^N Q_i(\rho_i - p),
\end{equation*}
is a closed loop optimal asset purchasing law that will drive the price of the asset along a path of least surprise (to observers). Any trader with sufficient power could enact such a trading strategy, thus showing that extreme inequality in market purchasing power can not only distort the market but can do so in an optimal way that minimises surprise; i.e., is optimally covert to all external observers of the price only.

\section{Conclusions and Future Directions}\label{sec:Conclusions}
In this paper, we constructed a model of a binary options market in which trader beliefs were modelled using tools from information geometry. At a fixed price, the dynamics of agent beliefs about that price were given by a Hamiltonian system. We showed that in the absence of external information, completely identical agent beliefs evolved along a $2N-2$ dimensional centre manifold with a one dimensional slow manifold and a two-dimensional stable manifold. Introduction of heterogeneity has the potential to decrease the dimension of the centre manifold, assuming certain sufficient conditions are met. We then investigated the impact unequal (powerful) agents can have on the market, showing that in markets with high levels of inequality, powerful agents can both move markets and engage in pump and dump strategies that not only manipulate price but also other agents beliefs. We concluded the paper by showing an optimal control law for market manipulation that generates a minimum of surprise for external observers by causing price to travel along an information geometric geodesic. 

There are several potential future directions of research. Generalising the results on fixed point stability and constructing a formal proof on the impact of unequal traders on the dynamics of the system is clearly an immediate extension. Additionally, in work outside the scope of this paper, we observed an interesting delay dynamic between the mean agent belief and the price for varying values of $\beta$. In particular, $\beta$ seems to introduce a delay between these two signals. To our knowledge, this is an emergent property of the system and models the time it takes for information (belief) to be factored into the market spot price. Exploring this property of the model would be of interest. Additionally, extending these results to more general market models (with more complex probability distributions) would be of interest, as would fitting real-world market data to these models.

\bibliographystyle{iopart-num}
\bibliography{main,Oscillator}

\end{document}